\documentclass[12pt,preprint]{aastex}
\usepackage{emulateapj5}
\newcommand{\mincir}{\raise
-2.truept\hbox{\rlap{\hbox{$\sim$}}\raise5.truept\hbox{$<$}\ }}
\newcommand{\magcir}{\raise
-2.truept\hbox{\rlap{\hbox{$\sim$}}\raise5.truept\hbox{$>$}\ }}
\newcommand{\minmag}{\raise
-2.truept\hbox{\rlap{\hbox{$<$}}\raise6.truept\hbox{$<$}\ }}
 
\usepackage{graphicx}

\def\arcdeg{\hbox{$^\circ$}}

\newcommand{\be}{\begin{equation}}
\newcommand{\ee}{\end{equation}}

\newenvironment{inlinefigure}{%
\def\@captype{inlinefigure}%
\noindent\begin{minipage}{\linewidth}\begin{center}}
{\end{center}\end{minipage}\smallskip}
\makeatother
\shorttitle{Constraining the CDM spectrum normalization}
\shortauthors{Basilakos \& Plionis}

\begin{document}


\title{Constraining the CDM spectrum normalization in flat dark 
energy cosmologies}

\author{Spyros Basilakos\altaffilmark{1,2}
Manolis Plionis\altaffilmark{1,3}}

\altaffiltext{1}{Institute of Astronomy \& Astrophysics, 
National Observatory of Athens, Palaia Penteli 152 36, Athens, Greece}
\altaffiltext{3}{Research Center for Astronomy \& Applied Mathematics,
Academy of Athens, Soranou Efessiou 4, GR-11527 Athens, Greece}
\altaffiltext{2}{Instituto Nacional de Astrof\'{\i}sica \'Optica y
Electr\'onica, AP 51 y 216, 72000, Puebla, Pue, M\'exico}

\begin{abstract} 
We study the relation between 
the rms mass fluctuations
on 8$h^{-1}$Mpc scales and $\Omega_{\rm m}$ using the 
recent clustering results of XMM-{\it
Newton} soft (0.5-2\,keV) X-ray sources, which have 
a median redshift of $z\sim 1.2$. The relation can be 
represented in the form $\sigma_{8}=0.34 (\pm 0.01) \Omega_{\rm
  m}^{-\gamma}$ where $\gamma\equiv \gamma(\Omega_{\rm m},w)$ and it is valid
for all $w<-1/3$ models. By combining the X-ray clustering and SNIa data
we find that the model which best reproduces the
observational data is that with:
$\Omega_{\rm m}\simeq 0.26$, $w\simeq -0.90$ and $\sigma_{8}\simeq
0.73$, which is in excellent agreement with 
the recent 3-year Wilkinson Microwave Anisotropy Probe results. 

\end{abstract}


\section{Introduction}
The combination of the recently acquired, high quality, 
observational data on
galaxy clustering, the SNIa Hubble relation
and the CMB fluctuations, strongly support a universe with flat
geometry and a currently accelerated expansion
due to the combination of a low 
matter density and a dark energy component 
(eg. Riess, et al. 1998; Perlmutter et al.
1999; Percival et al. 2002; Efstathiou et al. 2002; Spergel et al. 2003; 
Tonry et al. 2003; Schuecker et al. 2003; Riess et al. 2004; 
Tegmark et al. 2004; Seljak et al. 2004; Allen
et al. 2004; Basilakos \& Plionis 2005; Blake et al. 2006; 
Spergel et al. 2006;  Wilson, Chen \& Ratra 2006, for a review
see also Lahav \& Liddle 2006).

From the theoretical point of view various candidates  
of the exotic ``dark energy''
have been proposed, most of them described by an equation of state
$p_{Q}= w\rho_{Q}$ with $w<-1/3$ (see Peebles \& Ratra 2003 and
references therein). Note that a redshift dependence of $w$ 
is also possible but present measurements are 
not precise enough to allow
meaningful constraints (eg. Dicus \& Repko 2004; Wang \& Mukherjee
2006).  
From the observational point of view and for a flat geometry, 
a variety of studies indicate that 
$w< -0.8$ (eg. Tonry et al. 2003; Riess et al. 2004; 
Sanchez et al. 2006; Spergel et al. 2006; 
 Wang \& Mukherjee 2006 and references therein)

Another important cosmological parameter is the normalization of the
CDM power spectrum in the form of the rms density fluctuations
in spheres of radius 8$h^{-1}$Mpc, the so called $\sigma_{8}$. A 
tight relation between $\sigma_{8}$
and the $\Omega_{\rm m}$ has been  derived mainly using the cluster
abundance with $\sigma_{8}\simeq 0.52 \Omega_{\rm m}^{-0.52}$ for a $\Lambda$
cosmology (Eke, Cole \& Frenk 1996). 
Also, Wang \& Steinhardt (1998) generalizing to take into account
dark energy models (with $w\ge -1$) found:
$\sigma_{8} \simeq 0.5 \Omega_{\rm m}^{-0.21+0.22w-0.33\Omega_{\rm
     m}}$. 

In this letter we use the clustering of high-$z$ X-ray AGNs to
estimate a new normalization of the CDM 
spectrum, valid for spatially flat cosmological models and
also for $w\le -1$ (the so called Phantom models). Finally, combining
our results with SNIa data (Tonry et al. 2003), we put  
strong constraints on the value of the equation of state parameter. 

\section{X-ray AGN Clustering}
In a previous paper (Basilakos et al 2005)
we derived 
the angular correlation function of the soft (0.5-2\,keV) XMM 
X-ray sources using a shallow (2-10\,ksec) wide-field survey ($\sim
2.3$ deg$^{2}$).
A full description of the data reduction, source detection and flux 
estimation are presented in Georgakakis et al. (2004). Here we
describe only the basic points. The 
survey contains 432 point sources within an effective
area of $\sim 2.1$ deg$^{2}$ (for $f_x \ge 2.7 \times
10^{-14}$ erg cm$^{-2}$ s$^{-1}$ ), while for 
$f_x \ge 8.8 \times 10^{-15}$ erg cm$^{-2}$ s$^{-1}$ the effective
area of the survey is $\sim 1.8$ deg$^{2}$.
The details of the correlation function estimation, the
various biases that should be taken into account (the amplification
bias and integral constraint), the survey luminosity
and selection functions as well as issues related to possible stellar
contamination are presented in Basilakos et al. (2005). 
The redshift selection function of our X-ray sources was 
derived using the soft-band luminosity 
function of Miyaji, Hasinger \& Schmidt
(2000), and assuming the realistic luminosity dependent density 
evolution of X-ray AGNs and it predicts a characteristic depth of $z\simeq 1.2$ 
for our sample (for details see Basilakos et al. 2005).

Our aim here is to investigate the relation between the normalization
of the CDM power spectrum ($\sigma_8$) and $\Omega_m$ in flat
cosmologies with $w\le -1/3$. A through study of the theoretical clustering predictions from
different flat cosmological models to the actual observed angular
clustering of distant X-ray AGNs was presented in Basilakos \& Plionis (2005).
For the purpose of this study we use Limber's formula
which relates the angular, ${\rm w}(\theta)$, and the spatial, $\xi(r)$,
correlation functions. In the case of a spatially flat Universe,
Limber's equation can be written as:
\begin{equation}\label{eq:angu}
{\rm w}(\theta)=2\frac{H_{\circ}}{c} \int_{0}^{\infty} 
\left(\frac{1}{N}\frac{{\rm d}N}{{\rm d}z} \right)^{2}E(z){\rm d}z 
\int_{0}^{\infty} \xi(r,z) {\rm d}u
\end{equation} 
with $E(z)=[\Omega_{\rm m}(1+z)^{3}+(1-\Omega_{\rm
    m})(1+z)^{3(1+w)}]^{1/2}$. 
Also $r$ is the physical separation between two sources,  
having an angular separation, $\theta$, given by $r \simeq 
 (1+z)^{-1} \left(u^{2}+x^{2}\theta^{2} \right)^{1/2}$
(small angle approximation).
 The number of objects within 
a shell $(z,z+{\rm d}z)$ and in a given 
survey of solid angle $\Omega_{s}$ is: 
\be
\frac{{\rm d}N}{{\rm d}z}=\Omega_{s}
x^{2}(z) n_s \phi(x)\left(\frac{c}{H_{\circ}}\right) E^{-1}(z)\;\;.
\ee
where $n_s$ is the comoving number density at zero redshift and $x(z)$
is the coordinate distance 
\be
x(z)=\frac{c}{H_{0}} \int_{0}^{z} \frac{{\rm d}y}{E(y)}\;\; .
\ee
Finally, the selection function 
$\phi(x)$ (the probability 
that a source at a distance $x$ is detected in the survey)
is estimated by integrating the appropriate Miyaji et al. (2000)
luminosity function.

\subsection{The Evolution of Clustering}
It is well known 
(Kaiser 1984; Benson et al. 2000) that according to
linear biasing the correlation function of the mass-tracer 
($\xi_{\rm obj}$) and dark-matter one ($\xi_{\rm DM}$), are related by:
\be
\label{eq:spat}
\xi_{\rm obj}(r,z)=b^{2}(z) \xi_{\rm DM}(r,z) \;\;, 
\ee
where $b(z)$ is the bias evolution function.
Here we use the bias evolution model of Basilakos \& Plionis
(2001; 2003), where we also compared in detail
our model with that of Mo \& White (1996)
and Matarrese et al. (1997). As an example, our model
predicts an AGN bias which is by $\sim 30\%$ greater and
by $\sim 35\%$ lower 
than that of Mataresse et al. (1997) at $z=0$ and $z=3$,
respectively (see Basilakos et al. 2005).
We remind the reader that our bias model is based 
on linear perturbation theory and the 
Friedmann-Lemaitre solutions of the cosmological
field equations. For the case of a spatially flat cosmological model
our general bias evolution can be written as:
\be\label{eq:88} 
b(z)= {\cal A} E(z)+{\cal C} E(z) 
\int_{z}^{\infty}
\frac{(1+y)^{3}}{E^{3}(y)} {\rm d}y +1 \;\;.
\ee
Note that our model gives a family of bias
curves, due to the fact that it has two unknowns
(the integration constants ${\cal A},{\cal C}$). 
The value of ${\cal C}$ is approximately found to be 
$\simeq 0.004$, as was determined and tested 
in Basilakos \& Plionis (2003). Note that $E(0)=1$ and ${\cal A}=b_{0}-1-{\cal C}\int_{0}^{\infty}
\frac{(1+y)^{3}}{E^{3}(y)} {\rm d}y$, where $b_{0}$ is the bias at
the present time. 
We have tested the robustness of our results
by increasing ${\cal C}$  by a factor of 10 and 100 to find 
differences of only $\sim$ 5\% in the fitted 
values of $\Omega_{m}$ and $b_{0}$.
This behavior can be explained from the fact that the dominant term
in the right hand side of eq. (5) is the first term [$\propto
(1+z)^{3/2}$] while the second term has a slower dependence on redshift 
[$\propto (1+z)$].

We quantify the underlying matter distribution clustering 
by presenting the spatial correlation function of the mass 
$\xi_{\rm DM}(r,z)$ 
as the Fourier transform of the 
spatial power spectrum $P(k)$:
\be
\label{eq:spat1}
\xi_{\rm DM}(r,z)=\frac{(1+z)^{-(3+\epsilon)}}{2\pi^{2}}
\int_{0}^{\infty} k^{2}P(k) 
\frac{{\rm sin}(kr)}{kr}{\rm d}k \;\;,
\ee
where $k$ is the comoving wavenumber.
Note that the parameter 
$\epsilon$ parametrizes the type 
of clustering evolution (eg. de Zotti et al. 1990). 
In this work we utilize 
a clustering behavior which is constant in comoving 
coordinates ($\epsilon=-1.2$).

As for the power spectrum, we consider that of CDM models, 
where $P(k)=P_{0} k^{n}T^{2}(k)$ with
scale-invariant ($n=1$) primeval inflationary fluctuations
(we verified that a small change of $n$, eg. $n\simeq 0.95$ according to
the 3-year WMAP does not produce appreciable differences on our results).
In particular, we use the transfer function parameterization as in
Bardeen et al. (1986), with the corrections given approximately
by Sugiyama (1995) while the
normalization of the power spectrum is given by: 
\be 
P_{0}=2\pi^{2} \sigma_{8}^{2} \left[ \int_{0}^{\infty} T^{2}(k)
  k^{n+2} W^{2}(kR){\rm d}k \right]^{-1} \;\;. 
\ee
where $\sigma_{8}$ is the rms mass fluctuation
on $R=8h^{-1}$Mpc scales and $W(kR)$ is the window function given by 
\be 
W(kR)=\frac{3({\rm sin}kR-kR{\rm cos}kR)}{(kR)^{3}} \;\;.
\ee  
Note that we also use the
non-linear corrections introduced by Peacock \& Dodds (1994).  

\section{Cosmological Constraints}

\subsection{X-ray AGN Clustering likelihood}
Following the same notations as in Basilakos \& Plionis (2005) 
in order to constrain the cosmological parameters
we use a standard $\chi^{2}$ 
likelihood procedure to compare the measured 
XMM soft source angular correlation function (Basilakos et al. 2005) 
with the prediction of different spatially flat cosmological models.
The likelihood estimator\footnote{Likelihoods
  are normalized to their maximum values.} is defined as:
$${\cal L}^{\rm AGN}({\bf c})\propto {\rm exp}[-\chi^{2}_{\rm AGN}({\bf c})/2]$$
with:
\be
\chi^{2}_{\rm AGN}({\bf c})=\sum_{i=1}^{n} \left[ \frac{ {\rm w}_{\rm th}
(\theta_{i},{\bf c})-{\rm w}_{\rm obs}(\theta_{i}) }
{\sigma_{i}} \right]^{2} \;\;,
\ee 
where ${\bf c}$ is a vector containing the cosmological 
parameters that we want to fit and $\sigma_{i}$ is the uncertainty
of the observed angular correlation function. We make clear that 
we work within the framework of a flat cosmology 
with primordial adiabatic fluctuations and baryonic
density of $\Omega_{\rm b} h^{2}= 0.022 (\pm 0.002)$ 
(eg. Kirkman et al. 2003; Spergel et al. 2006). 
Also utilizing the results of the HST key project (Freedman et al. 2001)
we fix the Hubble constant to its nominal value; ie.,  
$h\simeq 0.72$, derived also by our previous AGN clustering 
analysis (Basilakos \& Plionis 2005).
In that work the 1$\sigma$
uncertainty of the marginalized value of $h$ was found to be 
only $\sim 0.03$. Note that since we fix in the following analysis
the values of both $h$ and $\Omega_b$, we do not take
into account their quite small uncertainties.

The corresponding vector that we have to fit
is ${\bf c} \equiv (\Omega_{\rm m},w,\sigma_{8},b_{0})$.
In this paper,    
we sample the various parameters as follows:
the matter density $\Omega_{\rm m} \in [0.01,1]$ in steps of
0.01; the equation of state parameter $w\in [-2,-0.35]$ in steps
of 0.05, the rms matter fluctuations $\sigma_{8} \in [0.4,1.4]$ in steps
of 0.02 and the X-ray sources bias at the present time
$b_{0} \in [0.5,3]$ in steps of 0.05.
Note that in order to investigate possible equations of state,
we have allowed the parameter $w$ to take values below -1. Such models 
correspond to the so called {\em phantom} cosmologies (eg. Caldwell 2002;
Corasaniti et al. 2004). 

The resulting best fit parameters are presented in 
Table 1.  
It is important to note that our estimate for the 
$\sigma_{8}$ parameter 
is in very good agreement with that derived ($\sigma_{8} \simeq 0.75$) 
by the recent 3-years WMAP results (Spergel et al. 2006).
In Fig.1 we present the 1$\sigma$, 2$\sigma$ and $3\sigma$
confidence levels in the $(\Omega_{\rm m},\sigma_{8})$, 
$(\sigma_{8},b_{0})$ and $(\Omega_{\rm m},b_{0})$, 
planes by marginalizing the first one over 
$b_{0}$ and w, the second
one over $\Omega_{\rm m}$ and w and the
latter over $\sigma_{8}$ and w; while the dot in Fig. 1 
corresponds to the best fitted values\footnote{Hereafter, when we marginalize over the 
equation of state parameter we will use $w=-1$.}.

\begin{inlinefigure}
\plotone{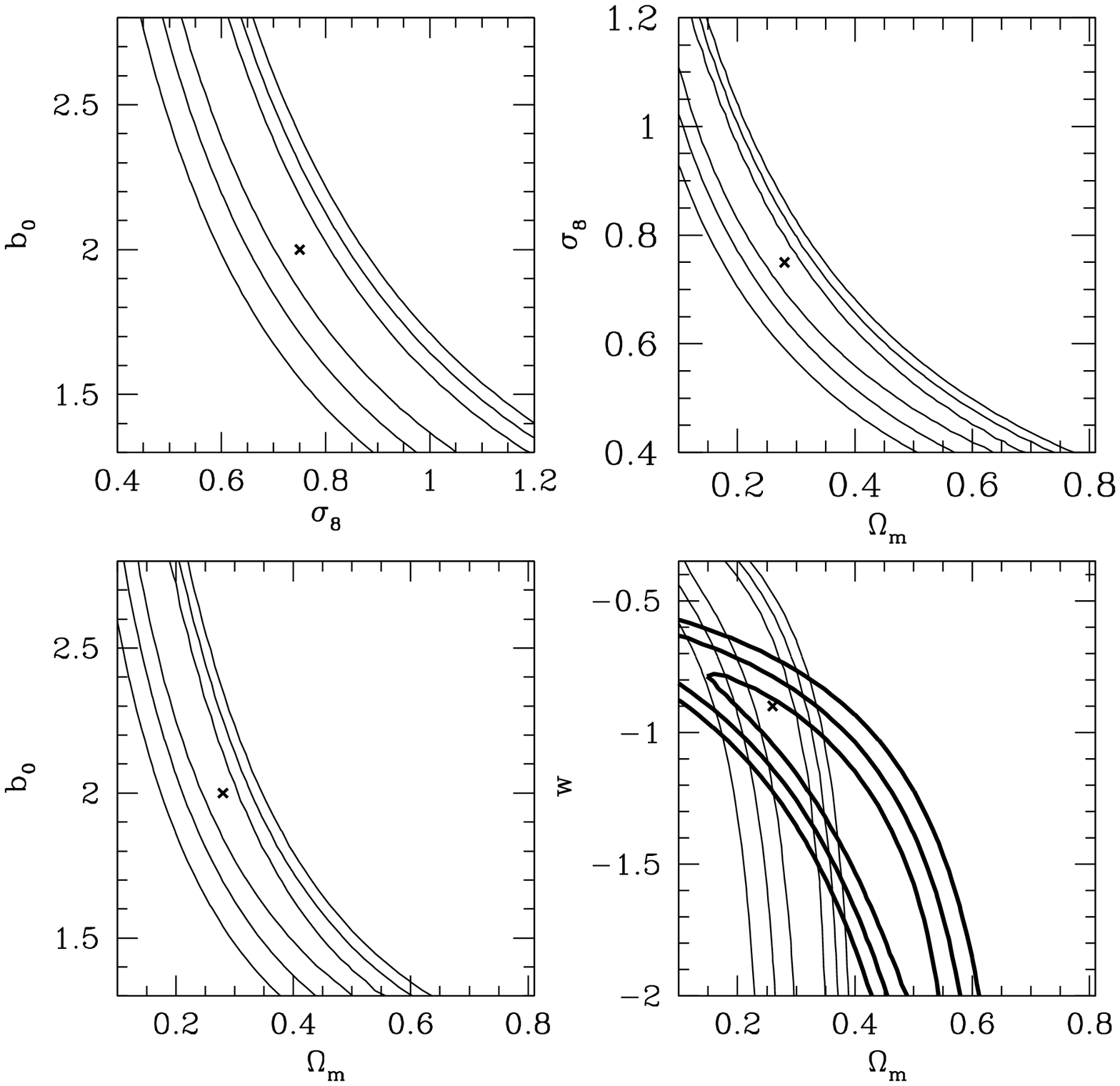}
\figcaption{Likelihood contours in the following planes: 
$(\Omega_{\rm m},\sigma_{8})$ (upper right panel), 
$(\sigma_{8},b_{0})$ (upper left panel), $(\Omega_{\rm m},b_{0})$ 
(bottom left panel) 
and the $(\Omega_{\rm m},{\rm w})$ (bottom right panel). The 
contours are 
plotted where $-2{\rm ln}{\cal L}/{\cal L}_{\rm max}$ is equal
to 2.30, 6.16 and 11.83, corresponding 
to 1$\sigma$, 2$\sigma$ and 3$\sigma$ confidence level. Finally, the thick
contours corresponds to the SNIa likelihoods.}
\end{inlinefigure}

Therefore, utilizing the clustering properties of our XMM sources 
and allowing for the first time values $w<-1$ 
(Phantom models) we derive the 
$(\Omega_{\rm m},\sigma_{8})$ relation, which can be fit
(within the $1\sigma$ uncertainty) by 
\be 
\sigma_{8}=0.34 (\pm 0.01) \; \Omega_{\rm m}^{-\gamma(\Omega_{\rm m},w)}
\ee
with
$$
\gamma(\Omega_{\rm m},w)=0.22 (\pm 0.04)-0.40 (\pm 0.05)w-0.052 (\pm
0.040)\Omega_{\rm m} \;\;.$$
In Figure 2 we present the results of the likelihood analysis for
different values of $w$ (points
with errorbars) and the previous fit as a continuous line.

Note that eq. (10) produces $\sigma_{8}$ values which are
significantly smaller than the usual cluster normalization (Wang \&
Steinhardt 1998) but are 
in good agreement with the 3-years WMAP results;  
for example for $w\simeq -1$ and $\Omega_{\rm m}\simeq 0.28$
we get $\sigma_{8}\simeq 0.73\pm 0.03$. 
It should be mentioned that in our previous work (Basilakos \& Plionis 2005) 
we had imposed a high $\sigma_{8}$ normalization, based on the 
cluster abundance, while here we leave the $\sigma_{8}$ parameter free.

The lower right panel of Fig. 1 shows 
the 1$\sigma$, 2$\sigma$ and 3$\sigma$
confidence levels (continuous lines) in the 
$(\Omega_{\rm m},w)$ plane by 
marginalizing over the $\sigma_{8}$ 
and the bias factor at the present time.
It is evident that $w$ is
degenerate with respect to $\Omega_{\rm m}$
and that all the values in the interval 
$-2\le w \le -0.35$ are acceptable
within the $1\sigma$ uncertainty. 
Therefore, in order to put further constraints on 
$w$ we additionally use a sample of 172 supernovae (see Tonry et al. 2003).

\subsection{The AGN+SNIa likelihoods}
Here we combine the X-ray AGN clustering properties 
with the SNIa data by performing a joined likelihood analysis and
marginalizing the X-ray clustering results 
over $\sigma_{8}$ and $b_{0}$
(see Table 1) and thus the vector ${\bf c}$ now becomes: 
${\bf c}\equiv (\Omega_{\rm m}, w)$. The SNIa 
likelihood function can be written as: 
$${\cal L}^{\rm SNIa}({\bf c})\propto 
{\rm exp}[-\chi^{2}_{\rm SNIa}({\bf c})/2]$$
with:
\be
\chi^{2}_{\rm SNIa}({\bf c})=\sum_{i=1}^{172} \left[ \frac{ {\rm log}
    D^{\rm th}_{\rm L}
(z_{i},{\bf c})-{\rm log}D^{\rm obs}_{\rm L}(z_{i}) }
{\sigma_{i}} \right]^{2} \;\;,
\ee 
where $D_{\rm L}(z)$ is the dimensionless luminosity
distance, $D_{\rm L}(z)=H_{\circ}(1+z)x(z)$
and $z_{i}$ is the observed redshift. 
The thick lines in Fig. 1 represents 
the $1\sigma$, $2\sigma$, and $3\sigma$,  
confidence levels in the $(\Omega_{\rm m}, w)$ plane.
We find that the best fit solution is $\Omega_{\rm m}=0.30 \pm 0.04$ 
for $w>-1$, in complete agreement with 
previous SNIa studies (Tonry et al. 2003; Riess et al. 2004).

We now join the likelihoods:
$${\cal L}^{\rm joint}(\Omega_{\rm m}, w)={\cal L}^{\rm AGN} \times 
{\cal L}^{\rm SNIa}\;\;,$$
which peaks at 
$\Omega_{\rm m}=0.26 \pm 0.04$ with $w=-0.90^{+0.10}_{-0.05}$.
Using eq. (10) we find that the normalization
of the power spectrum that corresponds to these cosmological
parameters is
$\sigma_{8}\simeq 0.73$. It should be pointed out that our 
results are 
in excellent agreement with those derived by 
Spergel et al. (2006) using the recent WMAP (3-years) data: 
$\Omega_{\rm m}\simeq 0.24$, $w\simeq -0.97$ and $\sigma_{8}\simeq 0.74$.

\begin{inlinefigure}
\plotone{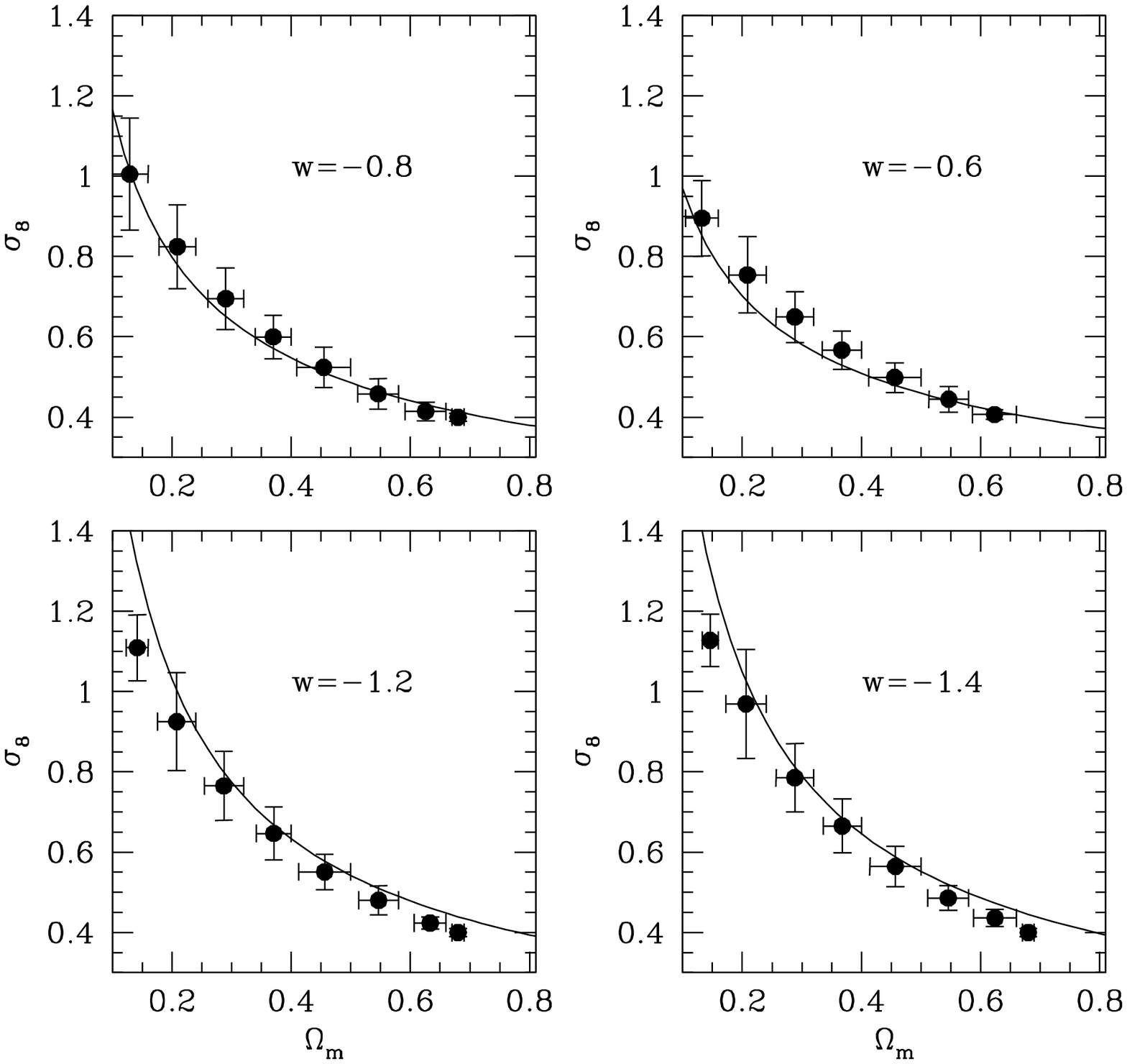}
\figcaption{The $(\Omega_{\rm m},\sigma_{8})$
  plane using different values for the equation of state parameter (points). 
The errors correspond to 1$\sigma$ (2.30) confidence levels. The
continuous line corresponds to the fit given by equation 10.}
\end{inlinefigure}

Many other recent analyzes utilizing different combinations of data
seem to agree with the former cosmological model.
For example, Sanchez et al. (2006) used the WMAP (1-year) CMB
anisotropies in combination with the 2dFGRS power spectrum and found 
$\Omega_{\rm m}\simeq 0.24$ and $w\simeq -0.85$, while Wang \&
Mukherjee (2006) utilizing the 3-years WMAP data together 
with SNIa and galaxy clustering results found $w\simeq
-0.9$ (see also Wilson et al. 2006). 

\section{Conclusions}
We have 
combined the clustering properties
of distant X-ray AGNs, identified as soft (0.5-2 keV) point sources in
a shallow $\sim$ 2.3 deg$^{2}$ XMM survey,
with the SNIa data. From the X-ray AGN clustering likelihood analysis alone
we have estimated the normalization of the CDM power spectrum and find
that the
rms density fluctuation in spheres of radius 8$h^{-1}$Mpc is fitted by:
$$
\sigma_{8}\simeq 0.34(\pm 0.01) 
\Omega_{\rm m}^{-0.22+0.40w+0.052\Omega_{\rm m}}
$$
which is valid also for Phantom models ($w<-1$).
Furthermore, a
joined likelihood analysis between the X-ray and SNIa data 
provides 
a best model fit with:
$\Omega_{\rm m}\simeq 0.26$ and $w\simeq -0.90$, which corresponds to
$\sigma_{8} \simeq 0.73$,
 in agreement with the recent 3-years WMAP results (Spergel et
al. 2006).

\begin{table}
{\small 
\caption[]{Cosmological parameters from the likelihood analysis:
The 1$^{st}$ column indicates the data used (the last row corresponds to
the joint likelihood analysis). Errors of the fitted parameters 
represent $1\sigma$ uncertainties. Note that for the joined
analysis the corresponding results are marginalized over 
the $\sigma_{8}$ and the bias factor at the present time, for which
we use the values indicated. 
}
\tabcolsep 10pt
\begin{tabular}{cccccc} 
\hline
Data& $\Omega_{\rm m}$& $w$& $\sigma_{8}$& $b_{0}$& $\chi^{2}/{\rm dof}$ \\ \hline 
{\rm XMM} &  $0.28\pm 0.03$  & uncons. ($w=-1$) & $0.75\pm 0.03$&  $2.0^{+0.20}_{-0.25}$ &0.90\\
{\rm XMM}$/{\rm SNIa}$& $0.26\pm 0.04$  &$-0.90^{+0.10}_{-0.05}$& $0.75$& $2.0$& 0.87\\ \hline 
\end{tabular}
}
\end{table}

\end{document}